\documentclass[aps,prl,twocolumn]{revtex4}
\newcommand{\beq}{\begin{equation}}
\newcommand{\eeq}{\end{equation}}
\newcommand{\beqa}{\begin{eqnarray}}
\newcommand{\eeqa}{\end{eqnarray}}
\newcommand{\beqar}{\begin{eqnarray*}}
\newcommand{\eeqar}{\end{eqnarray*}}

\def\ra{\rangle}

\def \D {{\Delta }}
\def \c {\otimes}

\def \la {\langle}
\def \ra {\rangle}

\def \s {\,\,\,\,}

\def \r {\rho}

\begin{document}

% Use the \preprint command to place your local institutional report
% number in the upper righthand corner of the title page in preprint mode.
% Multiple \preprint commands are allowed.
% Use the 'preprintnumbers' class option to override journal defaults
% to display numbers if necessary
%\preprint{}

%Title of paper
\title{A Thermodynamical Approach to Quantifying Quantum Correlations}

% repeat the \author .. \affiliation  etc. as needed
% \email, \thanks, \homepage, \altaffiliation all apply to the current
% author. Explanatory text should go in the []'s, actual e-mail
% address or url should go in the {}'s for \email and \homepage.
% Please use the appropriate macro foreach each type of information

% \affiliation command applies to all authors since the last
% \affiliation command. The \affiliation command should follow the
% other information
% \affiliation can be followed by \email, \homepage, \thanks as well.
\author{Jonathan Oppenheim$^{(1)(2)}$, Micha\l{} Horodecki$^{(2)}$, Pawe\l{}
Horodecki$^{(3)}$ and Ryszard Horodecki$^{(2)}$}
%04.70.D, 04.40, 05.70
%\homepage[]{Your web page}
%\thanks{}
%\altaffiliation{}

\affiliation{$^{(1)}$Theoretical Physics Institute, University of Alberta, 412
Avadh Bhatia Physics Laboratory, Alta., Canada, T6G 2J1.}
\affiliation{$^{(2)}$Institute of Theoretical Physics and Astrophysics,
University of Gda\'nsk, 80--952 Gda\'nsk, Poland}
\affiliation{$^{(3)}$Faculty of Applied Physics and Mathematics,
Technical University of Gda\'nsk, 80--952 Gda\'nsk, Poland}

%\email[]{jono@phys.ualberta.ca}
%Collaboration name if desired (requires use of superscriptaddress
%option in \documentclass). \noaffiliation is required (may also be
%used with the \author command).
%\collaboration can be followed by \email, \homepage, \thanks as well.
%\collaboration{}
%\noaffiliation

\date{\today}
%\maketitle
\begin{abstract}
We consider the amount of work which can be extracted from a heat bath
using a bipartite state $\rho$ shared by two parties.
In general it is less then the amount of work extractable when one
party is in possession of the entire state.  We derive
bounds for this "work deficit" and calculate it explicitly for a number
of different cases.
For pure states the work deficit is exactly
equal to the
distillable entanglement of the state, and this is
also achievable for maximally correlated states.
In these cases a form of complementarity exists between
physical work which can be extracted and
distillable entanglement.
The work deficit is a good measure of the quantum correlations in a
state and provides a new paradigm for understanding quantum non-locality.
\end{abstract}

% insert suggested PACS numbers in braces on next line
% insert suggested keywords - APS authors don't need to do this
%\keywords{}

%\maketitle must follow title, authors, abstract, \pacs, and \keywords
%\maketitle

%Early in the history of computer science it was realized that there
%is a
\maketitle
Strong connections exist between information, and thermodynamics.
Work is required to  erase a magnetic tape in an unknown state \cite{Landauer}
and bits of information can
be used to draw work from single heat bath \cite{szilard}\cite{Bennett-inf}.
The second law of thermodynamics forbids the drawing of work from
a single heat bath, however, if one has an engine which contains
"negentropy" (bits of information) then one can draw work from it.
The process does not violate the second law because the information is depleted
as entropy from the heat bath accumulates in the engine.
Typically, the source of
information is particles in a known states, and these states
can be thought of as a type of fuel or resource.
In particular,
quantum states can  be used as fuel\cite{Peres}, and recently,
physically realizable micro-engines have been proposed \cite{Scully}.

The field of quantum information theory has
also yielded tantalizing connections between entanglement and
thermodynamics \cite{forgetting}.  Bipartite states (jointly held by two
parties) such as the  maximally entangled state
\beq
\Psi_{AB}=\frac{1}{\sqrt{2}}\big(|00\ra+|11\ra \big),
 \label{eq:singlet}
\eeq
exhibit mysterious non-localities which can be exploited to perform
quantum useful logical work \cite{balance}
such as teleporting qubits \cite{teleportation}.  For many states, one can distill
singlets in order to perform logical work, but there is also {\it bound
entanglement} \cite{bound} which cannot be distilled from the state and it has
been proposed that this is analogous to heat \cite{therm}.  Pure bipartite states
can be reversible transformed into each other in a manner which is
reminiscent of a Carnot cycle \cite{RP,VP}. Furthermore, the
preparation of certain jointly held states
appears to result in a greater loss of information when
the state is prepared by two separated observers than when the entire state is
prepared by a single party \cite{nlwe}.  Connections between Landauer
erasure, measurement, error correction, and distillation have
also been explored \cite{Vedral-eras}.

In this paper we ask how much work can be drawn from a
single heat bath if the information is {\it distributed}
between two separated parties Alice and Bob.
It turns out that in general their engines will be
more efficient when information is localised, and that
the degree to which this is the case provides a powerful new
paradigm to understand and quantify non-locality in quantum mechanics.

As with the distant labs scenario for entanglement analysis,
we allow Alice and Bob to perform local operations on their
states, and communicate classically with each other (LOCC).
We will quantify the amount of potential work that cannot
be extracted by two separated parties by
introducing the concept of a {\it work deficit} $\D$,
defined to be the difference in the amount of work that can be
extracted from a state under LOCC versus the amount that can be
extracted by a party who holds the entire state.
For pure states 
%and  so called maximally correlated states  
we find
that the work deficit is exactly equal to the amount of distillable
entanglement $E_D$ of the state (i.e. the number of singlets that can be
extracted from the state under LOCC).  This also seems to be the
case for so called maximally correlated states. We also prove bounds for $\D$ and
show that it is a good measure for the amount of quantum
correlations present in a state jointly held by two parties.  
%Finally,
%a form of complementarity seems to exist between {\it physical} work and {\it logical}
%work, where distillable entanglement  is understood to be a
%source of "logical negentropy". 
A
more detailed analysis of the concepts introduced here will be presented
elsewhere \cite{inprep}.

Before proceeding with the quantum case it may be worthwhile to review
the connections between information and thermodynamical work for
classical states.  Consider a number of classical bits $n$ which
are all initially in the standard state 0.
%(for convenience, we will call this the $0$ state).
These bits can be used to
draw work from a heat reservoir of temperature $T$.  To visualize this, one
might imagine that a bit is represented by a box divided by a wall in the
center.  A particle placed in the left hand side represents the $0$ state, while
if the particle is in the right hand side, the bit is in the $1$ state.

Now imagine that we know that the bit is in the $0$ state.
We can draw work from the heat reservoir by replacing the central wall
with a piston and then allowing the particle to reversible push it out, drawing
$k T \ln 2$ of work from the reservoir
\cite{kT}.
We now no longer know where the particle is, so the entropy of the bit has been
increased by the same amount. No more work can be drawn from the bit, since we
don't know on which side to place the piston. 
%We could try to create a
%{\it perpetuum mobile} by measuring which side the particle is on, and then extracting
%more work from the bit, but the process of measuring the bit will
%increase the entropy of our memory, and erasing our memory costs at
%least one bit of work,
%a process known as Landauer erasure \cite{Landauer} (c.f.
%\cite{Bennett-inf,Lloyd}).

Although we cannot extract work from an unknown state, we can extract work
if we know that classical correlations exist.  Imagine for example
that we have two classical bits in unknown states, but
we know that they are in the same unknown state.  We can then perform a control
not (cnot) gate on the bits which flips the second bit if the first bit is a $1$.
After the cnot gate has been
performed the control bit is still in an unknown state, but the target bit is
now in the $0$ state and one bit of work can be extracted from it.  In general,
for a n-bit random variable $X$ with Shannon entropy $H(X)$ one can use the
first law of thermodynamics to see that the amount of work $W_C$ that can be
extracted is just the change in entropy of the state
\beq
W_C=n-H(X) \s .
\eeq

The same methods can be used to extract work from quantum
bits (qubits)\cite{Peres}.  If we have n
qubits in a state $\r$ and entropy $S(\r)$, then one can extract
\beq
W_t=n-S(\r) \s. \label{eq:wt}
\eeq
This is the amount of work that can be extracted in total by
someone who has access to the entire system $\r$.  As with the
classical case, all correlations can be exploited to extract work
from the state.

We now ask how much work two individuals can extract under LOCC using
a shared state $\r_{AB}$.
We imagine that Alice and Bob each have an engine which
can be used to locally extract work from a common heat bath.  Then, under
LOCC they try to extract the largest amount of local work $W_l$
possible. We then define the work deficit to be
the amount of potential work which cannot
be extracted under LOCC
\beq
\D\equiv W_t-W_l \s .
\eeq

Before proving some general results, it may be useful to give a few
simple examples. Consider the classically correlated
state \beq
\rho_{AB} = \frac{1}{2} \big( |00\ra\la 00| + |11\ra\la 11| \big)
\label{eq:simple}
\eeq
where the first bit is held by Alice, and the second by Bob.
We can see that $\D$ is zero for this case, since Alice can
measure her bit, send the result to Bob who can then extract
one bit of work by performing a cnot.  Alice can then reset the memory of her
measuring apparatus by using the work extracted from the bit that she held.
The total
amount of work extracted under LOCC is therefore $1$ bit, which
from Eq. (\ref{eq:wt}) is the same as
the amount of work that can be extracted under all operations.
In more detail, the steps involved in the process are: (a) Alice
uses a measuring device represented by a qubit prepared in the standard
state $|0\ra$.  She performs a cnot using her original state as the control
qubit, and the measuring qubit as the target.  (b) The measurement qubit
is now in the same state as her original bit and can be dephased (i.e.
decohered) in the $|0\ra,|1\ra$ basis so that the information is purely
"classical" (dephasing simply brings the off-diagonal elements of the density
matrix to zero, destroying all quantum coherence).  For this state,
dephasing doesn't change the state since
the state is already classical. (c) The
measuring qubit can now be sent to Bob who (d) performs a cnot using the
measuring qubit as the control. His original qubit is now in the standard state
$|0\ra$.
(e) Bob sends the measuring qubit back to Alice who (f) resets the measuring
device by performing a cnot using her original bit as the control.  Alice's
state is now in the same state as it was originally, while Bob's state is known
and can be used to extract $1$ bit of work.

We now consider how much work can be extracted from the maximally
entangled qubits of Eq. (\ref{eq:singlet}).
The same protocol as above can be used to find $\D=1$
(later we will show that this is optimal).  Alice and
Bob can extract one bit of work by following steps (a)-(f).
However, unlike the previous case, the measurement in step (b)
is an irreversible process and the original state and the entanglement
is destroyed by the dephasing that must occur for a measurement to be made.
On the other hand, someone with access to the entire state can extract two
bits of work since the state is pure and has zero entropy.

Basic questions now arise: How much work can be drawn from a given state
$\rho$? For which states is $\Delta=0$? How is $\Delta$ related to
entanglement?

To deal with these questions we have to state the paradigm
for drawing work from bipartite states more precisely. First
we will clarify the class of operations Alice and Bob are allowed
to perform. The crucial point is that here, unlike in usual
LOCC schemes, one must explicitely account for
all entropy transferred to measuring devices or ancillas.
So in defining the  class of allowable operations one must ensure that
no information loss is being hidden when operations
are being carried out.
One way to do this is to define elementary allowable operations
as follows:
(a) adding separable pure state ancillas to the system (b) local unitary
operations  (i.e. $U_A\c I_B$ or  $I_A\c U_B$ ) (c) sending qubits
through a dephasing channel.

The dephasing operation can be written as $\sum_i P_i \r_{AB} P_i$
where the $P_i$ are orthogonal local projection operators.
The class of operations (c) are
equivalent to local measurements and classical communication but has the
advantage that we don't need to worry about erasing the memory of measuring
devices.
Alice or Bob could also make measurements without sending the information,
but in this scheme it is wasteful.  Furthermore,
it is equivalent to Alice sending qubits down the dephasing channel and
having Bob send the qubits back.  We therefore do not need to consider
measurement and sending qubits as separate operations.

We imagine that Alice and Bob share an $n$ qubit state $\r_{AB}$ composed of
$n_A$ qubits held by Alice and $n_B$ by Bob.  The part of the state
which Alice holds is given by tracing out the degrees of freedom associated
with Bob's state i.e. $\r_A=Tr_B(\r_{AB})$ and visa-versa.  Alice and Bob then
perform combinations of operations (a)-(c) to arrive at a final state $\r_{AB}
'$ composed of $n'=n'_A+n_B'$ qubits. The difference $k=n'-n$ is the number of
pure state ancillas which they have added to the state, and therefore we must
subtract $k$ bits of work from the amount of work they can draw using the $n'$
bits. Since they must extract the work from $\r_{AB}$ locally, we find that
the total amount of local work that can be extracted is
\beqa
W_l&=&n'_A-S(\r'_A)+n'_B-S(\r'_B) - k \nonumber\\
&=& n - S(\r'_A)-S(\r'_B) \s.
\label{eq:workprec}
\eeqa
The goal is now clear.  Alice and Bob perform their operations to
arrive at a state which has
$S(\r'_A)+S(\r'_B)$ as low as possible.  They then draw work
locally using this new state $\r_{AB}'$.
%(if $n_{X}'=0$ then one says that $S(\rho_{X}')=0$, $X=A,B$).
%in fact then party $X$ has no $\rho_{X}'$).

Consider now the state of the form
\beq
\r_{AB}=\sum_{ij} p_{ij} | i_A\ra|j_B\ra\la j_B|\la i_A|
\label{eq:classical}
\eeq
where $| i_A\ra$ and $|j_B\ra$ are a local orthonormal basis.
Such states can be called {\it classically correlated} \cite{odroz}.
The natural protocol is that Alice sends her part to
Bob down the dephasing channel. This will not change the entropy of
the state. The final state $\r'$ will have $S(\r'_A)=0$ (strictly speaking
Alice will now have no system) while $S(\r'_B)=S(\r)$.
Thus according to Eq. (\ref{eq:workprec}) $W_l=n-S(\r)=W_t$, so that $\D=0$
for the above states. Note that
local dephasing in the eigenbases
of $\rho'_{A}$, $\rho_{B}'$ does
not change the optimal $W_{l}$ (\ref{eq:workprec})
but brings the state $\rho_{AB}'$ to the form (\ref{eq:classical}).
This gives another method to evaluate work:
instead of minimizing $S(\r'_A)+S(\r'_B)$ we can minimize
$S(\r')$ over classically correlated states $\r'$ that can be achieved
from $\r$ by the allowed class of operations.
Now we are in position to prove a general upper bound on the amount of work that
can be drawn using distributed information.

Our bound holds for pure states\cite{inprep}, but for more general states
our proof relies on the following 
%We shall prove the bound under the following 
assumption,
(although we conjecture that the bound holds in general).

{\it Assumption:} Bits which are sent down the communication channel are 
treated as classical in the sense that they are 
only dephased once, and not again in a second basis (which would destroy
the encoded information).

{\bf Theorem:}
{\it Under this assumption the maximum amount of work that can
be extracted using LOCC operations on an $n$ qubit state $\r_{AB}$
is bounded by $W_l\leq n - max\{S(\r_A),S(\r_B)\}$.}

The proof  follows after noting that Alice (or Bob), 
rather than directly sending the results of
measurements, can reversibly copy the measurement results 
by performing the cnot
operation with the measurement bit as the control bit and an ancilla as the
target. Alice can then send the copy to Bob who can use the information stored
in the copy. At the end of the whole protocol all the copies 
can be sent back (this follows from our assumtion)
and erased  by performing a second reversible cnot.
Alice and Bob's protocol will therefore not be more inefficient if they
keep their original measurement bits and only send copies to each other.
Consider now any optimal protocol transforming
$\rho_{AB}$ into the final state
$\rho'_{AB}$ of the form (\ref{eq:classical})
with minimal entropy $S(\rho'_{AB})$.
As we already know the protocol can be followed by dephasing
in the local eigenbases.
Hence before sending copies back and erasing them, the
entire system can be considered to be in another state $\sigma_{AB}$,
{\it still} in the form (\ref{eq:classical}), so
\cite{Wehrl}
$S(\sigma_{AB})\geq max\{S(\sigma_A),S(\sigma_B)\}$.
Now, as only copies of bit measurements were sent,
and the bits themselves were kept, production of
$\sigma_{AB}$ from $\rho_{AB}$ could only have increased
local entropies
%is invariant under
because neither unitary operation nor dephasing decreases entropy.
So one has
\begin{equation}
S(\sigma_{AB})\geq max\{S(\rho_A),S(\rho_B)\}
\end{equation}
where $S(\rho_A)$, $S(\rho_B)$ are local entropies
of the initial state $\rho_{AB}$.
Finally, because resending and erasing copies
preserves the spectrum of the whole state, one has
$S(\sigma_{AB})=S(\rho'_{AB})$ which gives
\begin{equation}
S(\r'_{AB}) \geq max\{S(\r_A),S(\r_B)\}.
\end{equation}

The theorem then follows directly from the fact that $n-S(\r'_{AB})$ is an upper
bound on the amount of work that can be drawn from the state $\r'_{AB}$.
The corresponding work deficit obtained under our assumption 
 %From the theorem it follows that under the assumption 
%about nondephasing measurement qubits) 
will be denoted by $\D_{r}$.
For mixed states it is possible that $ \D < \D_{r}$, but we conjecture
equality. 
%In fact, $\D_{r}= \D$ for pure states \cite{inprep}.
Also for 
one way LOCC schemes (classical communication from Alice to Bob only)
$\D_{r}$ coincides with the one way deficit $\D_{\leftarrow}$. 
From the theorem we have
\cite{lowerbound} 
\beq
\D_{r} \geq \max\{S(\r_A),S(\r_B)\} - S(\r)
\label{eq:bound}
\eeq
This allows one to calculate the extractable work for pure states by exhibiting
protocols that achieves this bound. To this end write a given pure state in
the Schmidt decomposition $\psi=\sum_ia_i|e_i\ra |f_i\ra$ where $e_i,f_i$ are
local bases. Alice then performs dephasing in her basis. The resulting state
is classically correlated and has entropy equal to $S(\r_A)$ where $\r_A$ is
the reduction of $|\psi\rangle \langle \psi|$. Note that the latter is the 
entanglement measure for pure states which is unique in the asymptotic regime and is equal to  the
distillable entanglement $E_D$ and the
entanglement cost (i.e. the number of singlets which are required to create the
state under LOCC) \cite{RP}. Thus for pure states the work deficit is
exactly equal to entanglement
\beq
\D(\psi)=E(\psi).
\eeq

We are able to calculate $\D_{r}$ for a broader class of states, the so called
maximally correlated states of the form
\beq
\r_{AB}=\sum_{ij}\sigma_{ij}|ii\ra\la jj|.
\label{eq:maxcor}
\eeq
 To achieve bound
(\ref{eq:bound}) Alice dephases her part in the basis $\{|i\ra\}$. The resulting
state is $\r'=\sum_i \sigma_{ii}|ii\ra\la ii|$, and has entropy equal to
$S(\r_A)$. Thus $\D_{r}(\r)=S(\r_A)-S(\r)$. One can check that for states
(\ref{eq:maxcor}) local entropies are equal and no smaller than the total
entropy so that $\D_{r}\geq 0$ as it should be. Now it turns out that for the above
states, we know $E_D$ \cite{Rains}, and it is again equal to $\D_r$. An example is
the mixture of state (\ref{eq:singlet}) with $1/\sqrt{2}(|00\ra-|11\ra)$: $E_D=
E_{D}^{\rightarrow}$ (the latter being one way distillable entanglement) 
is equal to $1-S(\r)$. The explicit distillation protocol attaining this value
was shown in \cite{huge}. That one can not do better follows from the relative
entropy bound \cite{VP,Rains} which is  equal to $1-S(\r)$ for those states.

The above result is rather surprising because the
state (\ref{eq:maxcor}) 
%is believed to 
contains bound entanglement 
i.e. the
entanglement cost of the state is greater than the entanglement
of distillation \cite{Vidal-recent}.
This result shows that 
%either 
work can be drawn from the bound entanglement.
% or the state
%does not contain bound entanglement, both situations being curious.

Although distillable entanglement cannot be used to perform physical work,
it allows us to perform {\it logical work}
(see \cite{balance}):  each bit of
distillable entanglement enables Alice to
teleport one qubit to Bob. For these states,
the total amount of extractable work $W_t$ gets divided between
physical work $W_l$ and logical work $E_D$.  Entanglement can therefore be
thought of as a source of non-local negentropy which can be used to
 perform logical work.  Just as with physical negentropy, logical negentropy
can only decrease under LOCC ($\delta\!E_D\leq 0$).  However, if
one uses the state to extract physical work the ability to perform logical work
is lost. Likewise, after performing logical work, the singlets are left in a
maximally entropic state and the ability to perform physical 
work is lost. There is thus a new form of complementarity 
between the logical and physical work.

%However the problem arises if one consider separable
%states - i.e. mixtures of product states \cite{Werner}. 
%%For one copy
%%it is easy to see that a state that is not of the form
%%(\ref{eq:classical}) has $\D\not=0$.
%Yet we do not know if, under collective actions, $\D$ will
%still be nonzero in the asymptotic limit of many copies. However, we
%conjecture that it is the case, and believe that an example might be some
%mixtures of pure states which exhibit
%so called {\it nonlocality without entanglement}
%\cite{nlwe}.
%%(note that asymptotic $\D\neq 0$ for separable states that
%%have entangled eigenvectors and a nondegenerate spectrum
%%has also not been excluded).
%(note that asymptotically, there appear to be seperable, locally
%distinguishable sets of states which also have $\D\neq0$).
%Thus we think that $\D$ will quantify not only entanglement, but
%rather nonlocality which is a broader notion.
%
%Finally it is interesting to compare our $\D$ with the measure  
%of quantum correlations proposed in \cite{Zurek-AnnPhys} called 
%{\it quantum discord}. It turns out that the discord (which depends 
%on choice of measurement) if maximized over measurements,
%is identical with our one-way work deficit. 
It is also worth investigating the connection between our approach and 
the measures of classical and quantum corelations introduced in 
\cite{VedralH} and \cite{Zurek-AnnPhys}.  
%It is also desirable to check the connections between
%$\D$ and the thermodynamical preparation-measurement irreversibility
%investigated in \cite{nlwe} 
%For example, there is a question how the quantity 
%$W_l-W_A-W_B\equiv W_l-n+S(\r_A)+S(\r_B)$  (the amount of classical 
%correlations that can be used to draw work by local engines) is 
%related to the measure of  classical correlations of Ref \cite{VedralH}.
It would also be desirable to  consider collective actions on many copies of the
given state. 
%It might be the case that such actions can decrease $\D$.
In the examples we considered, collective actions can't help since the
parameter $\Delta$ turns out to be additive.

In conclusion, we have proposed a paradigm for
quantifying quantum correlations 
motivated by thermodynamical and operational considerations.
%To our knowledge this is the first time 
%Quantum correlations are measured  
%thermodynamical  rather than by developing analogies. 
This approach is also fruitful in multipartite settings.  
The emerging function  $\Delta$ 
is nonzero for all entangled states, but need not 
vanish for separable states. It quantifies the 
part of correlations that must be destroyed during 
transmission via a classical channel (this is compatible with the 
observation that decoherence causes
a Maxwell demon to be less efficient \cite{qdemon}).
If a quantum channel 
was available, all information could be localized, and 
the full work $n-S(\r)$ could be drawn from local heat baths.
Thus the work deficit $\Delta$ quantifies truly quantum correlations. 
%Thus we think that $\D$ will quantify not only entanglement, but
%rather nonlocality which is a broader notion.
%It is conjectured to be upper bound  for distillable entanglement. 
Finally, we hope that the present  approach  will prove fruitful
in further investigations  of quantum properties of compound systems,
in particular that it may help discover a ``new face''  of the so called 
{\it thermodynamics of entanglement}.

\vspace{.2cm}
This work is supported by EU grant EQUIP, Contract No. IST-1999-11053.
J. O. thanks R. Laflamme and R.
Cleve for interesting discussions, and NSERC and DITP for support.

%\noindent { \small E-mail: jono@phys.ualberta.ca}

%\end{references}

\begin{thebibliography}{9}
\bibitem{kT}
Henceforth, we will set $T=k=1$ so that $k T \ln 2$ is simply "one
bit" of work.  We will also describe the dimension of Hilbert
spaces in terms of qubits.
\bibitem{odroz}
This is smaller class than separable states because the latter do
not need to have product eigenvectors.
\bibitem{lowerbound}
An upper bound for $\D$ is given by relative entropy
distance from the set of classically correlated states \cite{inprep}.
\end{thebibliography}
\end{document}